\documentstyle[twocolumn,prl,aps]{revtex}
\begin{document}
\title{Many-particle resonances in 
excited states of semiconductor quantum dots}
\author{Konstantin Kikoin and Yshai Avishai}
\address{Department of Physics, Ben-Gurion University, Beer-Sheva 84 105,
Israel}
\date{\today}
\maketitle

\begin{abstract}
Anderson impurity model for semiconductor quantum dot is extended to take
into account both particle and hole branches of charge excitations. It is
shown that in dots with even number of electrons where the Kondo effect is
absent in the ground state, novel midgap exciton states emerge in the
energy spectrum due to Kondo-type shake-up processes. The relevance of the
model to heterostructures doped by transition metal impurities 
and to rare earth ions adsorbed on metallic surfaces is discussed.
\end{abstract}



\section{Introduction}

In recent years, the physics of
single electron tunneling through quantum dots 
has been the focus of attention for 
experimentalists and theoreticians \cite{Rev91,Rev97}. In these 
mesoscopic devices 
the features of many-body quantum systems are manifested,
and modern technologies allow experimentalists to model various
realizations of strongly correlated electron systems, even those which
are hardly achievable in "natural" objects (metallic and semiconductor 
compounds). 

Within this context,
one of the most remarkable manifestations of collective quantum effects 
is resonance electron 
tunneling through quantum dots under the condition 
of strong Coulomb blockade. It was shown that
the physics of resonance tunneling 
can be formulated in terms of the Anderson impurity
model for strongly correlated electrons in metals
\cite{Glazr88}. It is established that the 
collective resonance occurs at 
the Fermi level, provided the 
Anderson impurity can virtually be regarded as a localized 
spin. This condition is fulfilled in quantum dots
when the strong Coulomb blockade 
fixes the number of electrons ${\cal N}$ in the dot 
to be an odd integer. Exchange 
correlations of the spins of electrons 
in the leads with the spin of 
the dot electron arise in electron cotunneling. 
This scenario is reminiscent of the
Kondo effect in metals doped by magnetic 
impurities. Indeed,
the Kondo features recently
predicted in Ref.\cite{Glazr88} 
were observed in the conductance of
quantum dots formed in GaAs/GaAlAs heterostructures \cite{Gogo98} and
Si-MOSFETs \cite{Rok99} by gate depletion techniques.

From the point of view of the 
conventional Anderson model, resonance 
scattering of the Kondo type as mentioned above is 
expected to be absent in even-${\cal N}$ quantum dots
with singlet ground state. 
Yet, we have recently demonstrated that 
this expectation is not quite justified \cite{PAK99}. 
A Kondo-type cotunneling can indeed
be induced by an external magnetic field 
in an even-${\cal N}$ quantum 
dot fabricated in two-dimensional electron gas 
(2DEG) which is usually formed 
in GaAs/GaAlAs or Si/SiO$_2$ heterostructures. 
This magnetically driven
Kondo tunneling arises in strong enough field when the Zeeman energy
compensates the energy of excitation from a ground singlet state to a
triplet exciton state.

In the present paper we consider 
the possibility of collective resonant effects
in another type of semiconductor quantum 
dots (SQD). These can be fabricated 
in a form of nearly spherical droplets of 
Si \cite{Tsu91}, 
InAs \cite{Banin96,Banin99},
InP \cite{Bert98}, 
CdSe \cite{Bawe96} and, presumably, other semiconductors. 
Similarly to bulk semiconductors,
such nanocrystals  have filled "valence
band" and empty "conduction band"
formed by spatially quantized levels. The energy gap dividing these two
groups of discrete levels is even wider 
than in their bulk parents, so it is natural to assume
that in the ground state of such a SQD ${\cal N}$ is even. 

Existence of electron and hole branches of charge excitations, as well as
exciton lines in optical spectra was demonstrated in
semiconductor nanocrystals. Moreover, evidences of Coulomb
blockade phenomena (which is a 
necessary precondition for collective
effects) are reported by some of 
the experiments cited above. 
Thus, the tunnel Hamiltonian for these 
quantum dots cannot be mapped 
on the standard Anderson model, since 
the existence of additional branches
of excitations in quantum dots 
should be taken into account. Indeed, discrete exciton 
states do not show up in a conventional situation of an
impurity immersed in a conduction band of a metal, even when such an impurity 
has internal charge degrees of freedom (e.g., rare-earth impurities with two
unfilled electron shells \cite{Kibur93}). However, long-living excitons can
still exist in SQD coupled with metallic leads, since 
the scattering center is spatially isolated from the conduction electrons. 

The question then arises, as to what kind of many-particle effects can be
observed in SQD with additional electron-hole 
branches of excitations. Confined excitons are created by light 
illumination, and possible collective effects can be observed only 
at finite frequencies in excited states of the system. 

Our goal in this paper is then to
modify the conventional theory of resonance tunneling
through quantum dots with odd ${\cal N}$ 
in the ground state for the case
of SQD with even ${\cal N}$ in the ground and excited states. 
It is demonstrated below that 
collective shake-up processes lead, 
under certain conditions, to formation of
a many-particle
midgap state, which can be observed, e.g., 
as an additional line in the
luminescence spectra. 
For this purpose we adapt, in section 2, the basic tunnel
Hamiltonian
of quantum dot \cite{Glazr88} to a more general case of SQD with
electron and hole branches of single-particle excitations. 
Then, in section 3 we develop a perturbation theory 
for optical line shapes in SQD and
demonstrate that new many-particle resonances arise in electron-hole
cotunneling through SQD with even ${\cal N}$ at finite frequencies. 
Optical response function of a semiconductor nanocrystal
is calculated and
finite lifetime effects are discussed, 
which significantly influence the
many-particle tunneling at finite frequency. Section 4 is devoted 
to the study of midgap excitons in SQD.
Since the ground state of an
even-${\cal N}$ quantum dot is a spin singlet, one should not expect radical
reconstruction of its spin structure due to multiple creation of
electron-hole pairs which is the key mechanism for the formation of a
Kondo-ground
state in the conventional theory. Besides, the finite life time of the
excited
state of SQD cuts effectively the infrared divergences in electron-hole
pair spectrum, and one can hope that a perturbation approach allows one to
encode the main features of the shake-up effect in the exciton spectrum
of semiconductor nanocrystals. 
Section 5 is dedicated to applications of the theory developed so far.
In particular, possibilities of realization of similar 
shake-up effects in other physical systems are discussed. The paper 
is then summarized in section 6.
\section{Model Hamiltonian}

The electronic structure of a quantum dot coupled to metallic leads is
usually described by some variant of the Anderson Hamiltonian
\begin{equation}
H=H_{d}+H_{b}+H_{t}=H_{0}+H_{t}.  \label{1.1}
\end{equation}
where $H_d$ is the Hamiltonian of an isolated dot which includes also pair
interactions between confined electrons, $H_b$ describes the band electrons
in the leads, and $H_t$ is the tunneling term which couples the electrons in
the dot with those in the leads (regarded as metallic reservoirs).

Unlike the situation encountered in quantum dots formed by external
electrostatic potential in 2DEG
confined in the boundary layer of semiconductor heterostructure
(HQD), the electronic spectrum of nanocrystal formed by the methods of
colloidal chemistry, ion implantation or strain-induced surface growth,
preserves, to some extent, the features of 
the parent (bulk) material. Our
choice of effective tunneling Hamiltonian 
is based on quantum-mechanical
calculations of the electronic spectra of real SQD (see, e.g. \cite{Zuf99}
and references therein), although we simplify these spectra and retain only
those features which are essential for the many-particle shake-up effects in
the exciton spectrum.

Another important difference from 
HQD, where one usually deals with a partially filled
conduction band, is a huge energy gap $E_g$ which
divides the excitation spectrum
from the ground (neutral) state of SQD. The magnitude of $E_g$ in
nanocrystals grows exponentially 
with decreasing radius of nanocrystal $r_n$, 
and substantially exceeds the corresponding 
gap in bulk semiconductors at, say, $r_n\sim 10-20$ \AA . 
Electron and hole eigenstates of SQD can be
classified in accordance with the angular 
symmetry of the nanocrystal, but the
Brillouin zone parentage of these 
states can be traced by projecting the dot
state $\psi _i$ onto the bulk states $\phi _{n{\bf k}},$
\begin{equation}
\psi _i=\sum_{n{\bf k}}C_{n{\bf k}}^i\phi _{n{\bf k}}  \label{1.0}
\end{equation}
(see ref. \cite{Zuf98}). Here the index $i=c\lambda ,v\lambda$ stands for
the
orbital quantum numbers $\lambda =lm$ of the discrete state of confined
electron (hole), $n$ and ${\bf k}$ are the band index and the wave vector of
the electron in a bulk semiconductor. Both electron and hole levels 
$\varepsilon _i$ are spatially quantized, and the distance $\delta
\varepsilon_i $ between neighboring levels is rather large. For
example, according to numerical calculations of Ref. \cite{Zuf99},
the first and second conduction levels in InP nanocrystal are derived
from the bulk $\Gamma_{6c}$ and $L_{6c}$ states, respectively. Both
levels are s-like, and the inter-level distance in a dot with
$r_n=10.11$ \AA~ is $\delta \varepsilon_c=0.343$ eV. Two first
valence states in this SQD are derived from the bulk $\Gamma _{8v}$
states, and both of them are doubly degenerate. Their envelope
functions are of s- and p-type respectively, and the inter-level
distance is $\delta \varepsilon _v=0.089$ eV. Experimentally
defined level spacings in InAs with $r_n=32$ \AA~ are 
$\delta\varepsilon _c=0.31$ eV and $\delta \varepsilon _v=0.10$ eV
\cite{Banin99}. These values substantially exceed the tunneling
amplitudes which couple the SQD with the leads.

As a consequence, when considering the lowest 
exciton states, restriction to 
the first conduction and valence levels $\varepsilon _{c,v}$ should be an
excellent approximation. Whence, the dot Hamiltonian reads,
\[
H_d=\varepsilon _cn_c-\varepsilon _vn_v+
\]
\begin{equation}
\frac{U_c}2n_c(n_c-1)+\frac{U_v}2n_v(n_v-1)+H_{cv},  
\label{1.2}
\end{equation}
where $\sigma $ is the spin projection (spin 1/2 is considered),
$n_c=\sum_\sigma d_{c\sigma }^{\dagger }d_{c\sigma }$ is the
occupation number for the electrons at an empty conduction 
level, $n_v=\sum_\sigma \overline{n}_{v\sigma }=\sum_\sigma
d_{v\sigma }d_{v\sigma }^{\dagger }$ is the occupation number for
holes at a filled valence level $\varepsilon _v$. We neglect the
actual four-fold degeneracy of the valence states, which 
is lifted in any case by Coulomb blockade. The charging
energy $U_{c,v}$ which is responsible for this effect is
estimated as $\approx 0.11$ eV in InAs nanocrystals
\cite{Banin99}, and it is substantially larger than the
tunneling width. The last term in eq. (\ref{1.2}) describes the
electron-hole interaction which lowers the energy of the
electron-hole pair and splits the exciton states into singlet
and triplet ones.

The band electrons in the left and right leads are labeled by the index
$\alpha
=L,R,$ respectively, thus the band Hamiltonian reads 
\begin{equation}
H_b=\sum_{k\sigma }\sum_{\alpha =L,R}\varepsilon _{k\alpha }c_{k\sigma
\alpha }^{\dagger }c_{k\sigma \alpha }.  
\label{1.3a}
\end{equation}
Here the sum over $k$ runs over the conduction band whose 
width $D$ is rather large (see below). The SQD is tuned in such a way
that the Fermi level $\varepsilon _F$ 
(which is taken as the reference zero energy) falls into the
energy gap of the quantum dot, $E_g=E_c-E_v$. Finally, the
tunneling term in the Hamiltonian (\ref{1.1}) has the usual form
\begin{equation}
H_t=\sum_{k\sigma \alpha }\sum_{j=c,v}(V_{j}^\alpha c_{k\sigma \alpha
}^{\dagger }d_{j\sigma }+\text{h.c.}),  
\label{1.3b}
\end{equation}
where $V_{(c,v)}^\alpha $ are the corresponding tunneling amplitudes
(we neglect the $k$-dependence of these integrals). 
In the strong interaction regime on which we focus 
our attention, the tunnel width
$\Gamma_i^\alpha=\pi \rho _0|V_i^\alpha|^2$ is, in general, rather small
and 
different charge states of the SQD are not mixed by tunneling. 

Having thus specified the precise form of the Hamiltonian
it is useful at this point to recall
the hierarchy of energy parameters in SQD, that is,
\begin{equation}
D, E_g >\delta \epsilon _i>U\gg \Gamma^\alpha_i .  \label{1.3c}
\end{equation}
With these data at hand, 
anticipating the use of perturbation
theory for strongly correlated electrons \cite{Gun83}, 
it is useful to rewrite the reduced 
dot Hamiltonian $H_d$ in a diagonal form by means of Hubbard projection
operators $X^{\Lambda \Lambda ^{\prime }}\equiv |\Lambda \rangle \langle
\Lambda ^{\prime }|$,
\begin{equation}
H_d=\sum_\Lambda E_\Lambda X^{\Lambda \Lambda },  
\label{1.3d}
\end{equation}
where $E_\Lambda $ are the energy levels of the dot corresponding
to
different occupation numbers $\{n_c,n_v\}$. The lowest levels are
$E_{1,0}=E_c$, $E_{0,1}=E_v$, $E_{1,1}=\Delta$, $E_{2,0}=2\Delta_c+U$,
$E_{0,2}=2\Delta_v+U$.
Here $\Delta_{c}=\varepsilon _c-\varepsilon _F$, 
$\Delta_v=\varepsilon _F-\varepsilon
_v$, $\Delta =E_g-U_{cv}$ is the energy of the lowest exciton state,
which is
obtained after discarding the exchange splitting $\sim$ several meV
\cite{Zuf99} between the singlet and
triplet
states. The electron-hole attractive interaction is estimated as
$U_{cv}\approx 0.2$ eV in InP nanocrystals \cite{Zuf99,Zuf98}.

The tunnel operator $H_t$  changes
the number of electrons in a quantum
dot, hence it is convenient to rewrite
it in terms of projection operators $X^{\Lambda\Lambda'},$
but first we rationalize it by making a transformation to
standing wave basis \cite{Glazr88,PAK99}. Droping the ubiquitous $k$ 
dependence the transformation reads,
\begin{eqnarray}
c_{j\sigma +} &=&u_jc_{\sigma L}+ w_jc_{\sigma R}\nonumber \\ 
c_{j\sigma-} &=&w_jc_{\sigma R}-u_jc_{\sigma L} 
\label{1.6}
\end{eqnarray}
Here $u_j=V_{j}^{L}/V_j,\;w_j=V_{j}^{R}/V_j,\;
V_j=\protect\sqrt{|V_{j}^{L}|^2+|V_{j}^{R}|^2}.$ 
Only the operators $c_{kj\sigma +}$ enter the tunnel Hamiltonian 
which now assumes the simple form
\begin{equation}
H_t=\sum_{k\sigma \alpha }\sum_{j=c,v}(V_jc_{kj\sigma +}^{\dagger }d_{j\sigma
}+\text{H.c.})
\label{1.7}
\end{equation}
Hereafter it is assumed that the SQD coupling to the leads is
symmetric, that is, $V_{j}^{L}=V_{j}^{R}=V_{j}$,
$c_{kj\sigma}=c_{k\sigma}$.  
Now, starting with
the "vacuum" state $|0\rangle$ of quantum dot in a configuration
$\{0,0\}$ and filled Fermi sea, we find that $H_t$ connects this state
with a singlet exciton state $|es\rangle=\{1\sigma,1\sigma\}$ via the
intermediate states $|kv\rangle$, $|kc\rangle$  with one electron
or one hole in the dot respectively. These normalized states
can be introduced as follows,  
\begin{equation}
|kc\rangle =\frac{1}{\sqrt{2}}\sum_{\sigma }d_{c\sigma }^{\dagger}
c_{k\sigma }|0\rangle ,\;\;
|kv\rangle =\frac{1}{\sqrt{2}}%
\sum_{\sigma}c_{k\sigma }^{\dagger } d_{v\sigma }|0\rangle ,
\label{1.8}
\end{equation}
\begin{equation}
\left| es\right\rangle =\frac{1}{\sqrt{2}}\sum_{\sigma }d_{c\sigma }^{\dagger
}d_{v\sigma }|0\rangle \equiv B^{\dagger }\left| 0\right\rangle .
\label{1.2a}
\end{equation} 
The tunnel Hamiltonian can then be expressed
via projection operators
\begin{eqnarray}
H_t=V_c\sum_k\left(|kc\rangle\langle 0|
+\frac{1}{\sqrt{2}}|kv\rangle\langle es|\right) +H.c. \nonumber\\
+ V_v\sum_k \left(|kv\rangle\langle 0|
-\frac{1}{\sqrt{2}}|kc\rangle\langle es|\right) + H.c.\;, 
\label{1.2f}
\end{eqnarray}
with matrix elements given by
\begin{eqnarray}
\langle 0|H_{t}|kc\rangle = \sqrt{2}\langle kv|H_{t}|es\rangle =
V_{c}\nonumber \\
\langle 0|H_{t}|kc\rangle = -\sqrt{2}\langle
kc|H_{t}|es\rangle = V_{v}^{\ast }.
\label{1.2d}
\end{eqnarray}
It is worth noting that the triplet exciton states
\begin{equation}
|et,\sigma\rangle=d_{c\sigma }^{\dagger}d_{v,-\sigma }|0\rangle,\;\;
|et,0\rangle=\frac{1}{\sqrt{2}}\sum_{\sigma }\sigma d_{c\sigma }^{\dagger
}d_{v\sigma }|0\rangle
\label{1.2c}
\end{equation}
are coupled to the above states only in higher orders of perturbation
theory.

\section{Perturbation theory for optical line shape}
							
In this section the basic equations of perturbation
theory modified for excited states of SQD are derived. 
Evidently, no Kondo coupling is expected in
the ground state $\Psi_G$, and the strong Hubbard repulsion $U$
suppresses double occupation of electron or hole levels. Hence,
only states with singly charged dot are admixed by tunneling
to the neutral state $|0\rangle$ of the isolated dot. One can
consider this admixture as a small second order perturbation,
which results in a trivial shift of the ground state energy
$\delta E_G \sim -\Gamma \ln (D/E_g)$. We will see, however, that
Kondo-like processes do develop in the spectrum of {\it electron-hole
excitations} of the quantum dot given by the operator
$B^\dagger=|es\rangle\langle 0|.$ These states can be excited by
means of photon absorption process, which is described by the
Hamiltonian
\begin{equation}
H^{\prime}= \sum_{i\neq j}\sum_{\sigma}P_{ij}d_{i\sigma
}^{\dagger }d_{j\sigma }\exp(-i\omega t)+ h.c. 
\label{2.0}
\end{equation}
Here $P_{ij}$ is the matrix element of the dipole operator $\hat{P}$.

The optical line shape at photon energy $h\nu$ is given by the
Kubo-Greenwood formula ,
\begin{equation}
W(h\nu )\simeq {\rm Im}
\frac{1}{\pi }\langle s|\hat{P}\hat{R}(h\nu )\hat{P}| s\rangle
\label{2.1}
\end{equation}
where $\hat{R}(z)=(z - H)^{-1}$, and $\langle s|\ldots| s\rangle$ means
averaging the initial state over the equilibrium ensemble. We use the
temperature
perturbation theory based on the Brillouin-Wigner expansion for the
resolvent
\begin{equation}
R(z)=(z-H)^{-1}=R^{(0)}\sum_{n=0}^\infty [H_t R^{(0)}]^n,
\label{2.2}
\end{equation}
where $H_0=H-H_t$, $R^{(0)}=(z-H_0)^{-1}$ (see, e.g. \cite{Gun83}). This
expansion should be
inserted in the partition function ${\cal Z}$, which can be written in the
form
\begin{equation}
{\cal Z}={\rm Tr_dTr_b}\{e^{-\beta H}\}= \frac{1}{2\pi i}\int_C e^{-\beta z}%
{\rm Tr_dTr_b}(z-H)^{-1} dz.
\label{2.3}
\end{equation}
Here the trace is taken over the dot and band states of $H_0$,
 and the
contour $C$ encircles all the singularities of the integrand. 
Carrying out
the summation over the states $E_b$ of the conduction
electrons we can obtain a system of equations for the diagonal
matrix elements of the reduced resolvent $R_{\Lambda\Lambda}(z)$,
\begin{equation}
R_{\Lambda\Lambda}(z)=\frac{1}{{\cal Z}_b}\sum_b
\frac{e^{-\beta E_b}}{z-E_\Lambda} \langle\Lambda|\langle b|
\sum_{n=0}^\infty [H_t R^{(0)}]^n
|b\rangle |\Lambda\rangle ,
\label{2.9}
\end{equation}
projected on the low-lying states
$|\Lambda\rangle=|0\rangle,|es\rangle$ of the neutral quantum dot
$({\cal Z}_b$ is the partition function of the lead electrons). As
a result, the problem of line-shape function (\ref{2.1}) reduces
to calculation of the retarded exciton Green function
\begin{equation}
G^{R}_{ee}(z) =-i\int dte^{izt}\theta(t)\langle 
| [B^\dagger(t) B(0)]|\rangle .
\label{2.5a}
\end{equation}

Coulomb blockade and other restrictions
imposed by the system of inequalities (\ref{1.3c}) eliminate 
the possible
admixture of excited states with more than one electron (hole) in
the SQD. The use of standard Feynmann diagram
technique is therefore impractical. 
However, partial summation of perturbation series is still
possible employing methods of Laplace
transform (\ref{2.5a}) \cite{Gun83,Bickers}. Calculations to lowest
order involve only the states $|kc\rangle$ and
$|kv\rangle$ (\ref{1.8}) as intermediate states in the
perturbation series (\ref{2.9}) which are admixed by the tunnel
operators (\ref{1.2d}) to the ground and exciton states
$|\Lambda\rangle=|0\rangle, |es\rangle$. This approximation
neglects intermediate states with multiple electron-hole pairs
in the leads (see, e.g., \cite{Lacr81}). 
In conventional theory, it is valid at high
temperatures $T>T_K$. The
response function (\ref{2.9}) can 
then be found from the solution of the
system of equations for the matrix elements 
$R_{\Lambda'\Lambda}(z)=\langle\Lambda'|(z-H)^{-1}|\Lambda\rangle$,
which, in this
approximation, has the following simple form
\begin{eqnarray}
R_{ee}=R_{e}^{(0)}\left(1+\Sigma_{ee}R_{ee}+\Sigma_{e0}R_{0e}
\right)\nonumber\\
R_{0e}=R_{0}^{(0)}\left(\Sigma_{00}R_{0e}+\Sigma_{ee}R_{ee}
\right).
\label{2.5b}
\end{eqnarray}
The structure of these equations is illustrated in Fig.1.
Here
$
R_\Lambda^{(0)}=\langle \Lambda|(z-H_0)^{-1}|\Lambda\rangle=
(z-E_\Lambda)^{-1}
$
are the zero order matrix elements of the resolvent.
The self energies are given by
\begin{eqnarray}
\Sigma_{00} & = & \Sigma_{cc}^{+}+\Sigma_{vv}^{-},\;\; 
\Sigma_{ee} = \left(\Sigma_{vv}^{+}+\Sigma_{cc}^{-}\right)/2,  \nonumber \\
\Sigma _{0e} & = & \left(\Sigma_{cv}^{-} -\Sigma_{vc}^{+}\right)/\sqrt{2},  
\label{1.13}
\end{eqnarray}
where
\begin{equation}
\Sigma_{jl}^{+}= \sum_{k}\frac{V^*_jV_lf_k}{z -E_{c}+\varepsilon _{k}},\;
\Sigma_{jl}^{-}= \sum_{k}\frac{V^*_jV_l\bar{f}_k}{z +E_{v}-\varepsilon _{k}},
\label{sigmaP}
\end{equation}
in which $f_k$ is the equilibrium distribution function 
for lead electrons and $\bar{f}_k=1-f_k$.  
The exciton Green function can now be extracted from equation (\ref{2.5b}) 
\begin{equation}
R_{ee}(\epsilon )=\frac{z -\Sigma_{00}(\epsilon)}
{{\cal D}(\epsilon )}.
\label{2.9a}
\end{equation}
The poles of this function are determined by
the
equation
\begin{equation}
{\cal D}(\epsilon))=\det \left|
\begin{tabular}{cc}
$\epsilon -\Sigma_{00}(\epsilon)$ & $\quad -\Sigma _{e0}(\epsilon)$ \\
$-\Sigma_{0e}(\epsilon)\quad $ & $\epsilon -\Delta -\Sigma
_{ee}(\epsilon)$%
\end{tabular}
\right| =0.
\label{2.10}
\end{equation}
The real parts of the self energies 
\begin{equation}
{\rm Re}\Sigma^+_{jl}(\epsilon)\approx \frac{\Gamma_{jl}}{2\pi }\ln 
\frac{(\epsilon
-\Delta_{c})^2+(\pi T)^2}{D^2},  
\label{sigmaC}
\end{equation}
\begin{equation}
{\rm Re}\Sigma^-_{jl}(\epsilon)\approx \frac{\Gamma_{jl}}{2\pi }\ln 
\frac{(\epsilon
-\Delta_{v})^2+(\pi T)^2}{D^2},  
\label{sigmaV}
\end{equation}
have sharp maxima at  
energies $\Delta_c$ and 
$\Delta_v$ respectively
$(z=\epsilon+is$, 
$\Gamma_{jl}=\pi \rho _0 V_j^*V_l).$
These peaks which are related to 
logarithmic singularities
result in novel
features in the exciton spectra (see next section). It should be
emphasized, however, that the lowest order approximation is not sufficient 
even for qualitative description of exciton states, because the electron-hole
pair has finite lifetime. Even when the conventional mechanisms of 
electron-hole recombination in the dot are ineffective at the Kondo time scale,
the tunnel width of confined electron and hole should be taken into account.

The processes which are involved in 
the formation of exciton self energies
up to the 4$^{th}$ order are displayed 
graphically in Figs. 2,3.  The diagram  
rules are obtained by straightforward generalization of the
corresponding rules described by Keiter and Kimball
(see Refs. \cite{Gun83,Bickers}). There are four 
types of vertices (Fig. 2) which couple 
the ground and exciton states of the dot  
with electron-hole states (see eq. \ref{1.2d}). The wavy lines stand 
for the dot propagator $R^{(0)}_{e,0}$, the solid and dashed lines
correspond 
to the charged (electron and hole) states of the dot and excess electrons 
and holes in the leads respectively. The Fermi factors $f(\varepsilon)$ and 
$\bar{f}(\varepsilon)=1- f(\varepsilon)$ are 
respectively assigned to the conduction
electron and hole lines. In the energy denominator 
$(z-E_\alpha)^{-1}$ corresponding to each vertical crossection of the 
block between two adjacent tunnel vertices, $E_\alpha$ is a sum of
energies $\pm\varepsilon_i$ and $\pm\varepsilon_k$ with the sign  $\pm$ 
respectively assigned to electron and hole propagators. 
  
The diagrams $(a,b)$ presented in Fig. 3 are taken 
into account in eq. (\ref{2.9a}) for $R_{ee}$. Each electron-hole
loop in these diagrams contains 
a logarithmic singularity (\ref{sigmaC}) or
(\ref{sigmaV}). The double wavy lines in the 4$^{th}$-order diagrams 
$(b)$ stand for 
the Green's function $G_{00}$ containing the 
electron-hole loops $\Sigma_{00}$ and $\Sigma_{0e}$ (eq. \ref{1.13})
in its self energy. 
In lowest order, exciton 
finite lifetime effects can be included in the theory within
the non-crossing approximation (NCA) \cite{Gun83,Bickers}. 
Such diagrams appear in 4$^{th}$ order of perturbation
theory (diagrams $(c)$ in Fig.3). 
It should be noted that in these diagrams, 
intermediate states include also 
triplet excitons. 

Within the NCA, the
internal propagators in the self energies 
(\ref{sigmaP}) can be ``dressed''. 
The corresponding integrals are then modified 
as follows,  
\begin{eqnarray}
\widetilde{\Sigma}_{jl} ^{+}(\epsilon) & = & \frac{\Gamma_{jl}}{\pi }
\int_{-D}^D \frac{f(\varepsilon)d\varepsilon}{\epsilon-E_{c}+\varepsilon-
B^{+}_{jl}(\epsilon+\varepsilon)},  \nonumber \\
\widetilde{\Sigma}_{jl}^{-}(\epsilon) & = & \frac{\Gamma_{jl}}{\pi }
\int^{D}_{-D} \frac{\bar{f}(\varepsilon)d\varepsilon}{\epsilon+E_{v}-%
\varepsilon- B_{jl}^{-}(\epsilon-\varepsilon)},  
\label{1.15}
\end{eqnarray}
where $B_{jl}^{\pm}$ are the integrals obtained through
insertions of electron (hole) lines in the self energies 
of the dot.
These diagrams describe  
damping of electrons and holes 
in the intermediate charged states of the
dot due to multiple 
electron-hole pair creation in the leads. In the next section 
corrections related to this damping mechanism
are discussed in some details. 

\section{Midgap excitons in semiconductor quantum dots} 

It might be instructive to
start the analysis of exciton states in even-${\cal N}$ SQD 
by inspecting an extreme limit of completely localized 
hole whose wave function does not overlap with those of the lead
electron states, that is, $V_v=0$.  
Tunneling is possible in this case  only to the
empty state $\varepsilon_c$ above the Fermi level. Consequently,
the self energies $\Sigma_{\Lambda\Lambda'}$ (\ref{1.13}) contain only
the contributions $\Sigma_{cc}^\pm$.  
In the conventional situation \cite{Gun83} 
this case corresponds
to a trivial singlet ground state with $n_c=0$ 
where states with $n_{c\sigma}=1$
are decoupled from the low-energy excitations. The second order correction 
$\Sigma_{cc}^+$ merely results in a trivial 
shift $\delta E_0$ of the 
{\it ground state} energy (see above). However, in SQD the
corresponding correction 
$\Sigma_{cc}^-$ to the {\it exciton energy} $\epsilon=\Delta$ has a 
logarithmic singularity at an energy $\Delta_v$ 
(see eq. \ref{sigmaV}), and the 
weak coupling perturbation theory becomes invalid at some characteristic
"Kondo" temperature $T_K=D\exp (-2\pi\Delta_v/\Gamma_c)$. At $T>T_K$ 
this correction results in 
an additional peak in the exciton spectral 
function $\sim {\rm Im} R_{ee}(\epsilon)$ at $\epsilon\approx \Delta_v$.   
Possible physical realizations related to this new 
type of solution are discussed in the
concluding  section. 

Next, let us consider
a more realistic (and complicated) situation when the 
tunneling is allowed both in electron
and hole channels, but the hole state is out of
resonance with the band continuum. It
arises, e.g., in a device whose 
leads are prepared from a strongly doped 
$n$-type semiconductor with a gap located against the top of the 
valence band of the SQD (see Fig. 4).
The pairs GaAs(lead)/InAs(dot) or CdSe(lead)/ZnTe(dot) 
are possible candidates for such heterostructures.
One may also tune the level $\varepsilon_v$ in the energy gap of the 
lead semiconductor by applying an external gate voltage.
In principle, structures with damped hole and undamped  electrons
in the dot can also be fabricated. 

Consider now an asymmetric configuration with  
strongly localized hole, $\Delta_v >\Delta_c$, and assume that
\begin{equation}
\Delta_v - \Delta_c\gg \Gamma_{c,v}\;\;\; 
\eta\equiv\left(\frac{V_v}{V_c}\right)^2\ll 1\;.  
\label{3.0}
\end{equation}
The second inequality reflects the known fact that hole states tend 
to be more confined inside the dot and more compact 
than electron states
\cite{Zuk98}. Again, in lowest order  
of perturbation theory which neglects intermediate states with 
electron-hole pairs in the leads the spectral density given by 
Im$R_{ee}(\epsilon)$ displays additional peaks along with the conventional 
exciton peak at $\epsilon\sim \Delta$. The latter can be obtained from 
eqs (\ref{2.9a}) and (\ref{2.10}) by representing the secular equation
in a form
\begin{equation}
\epsilon\approx \Delta+\Sigma_{ee}(\Delta)+
\frac{|\Sigma_{0e}(\Delta)|^2}{\Delta}.
\label{3.1}
\end{equation}
Note that the logarithms in the self 
energies (\ref{sigmaC}),(\ref{sigmaV}) are
small at  $\epsilon\sim \Delta$, so this equation merely describes a weak 
renormalization of the bare exciton state $\epsilon = \Delta$ caused by
virtual tunneling processes.

Novel aspects are exposed in the behavior of $R_{ee}$ 
(see equation \ref{2.9a}) at $\epsilon\sim \Delta_v$
where $\Sigma^-_{ij}$ have a sharp maximum. The secular equation 
(\ref{2.10}) may then be rewritten as
\begin{equation}
2(\epsilon -\Delta) = \Sigma_{vv} ^{+}+\Sigma_{cc} ^{-}+ 
\frac {|\Sigma_{cv}
^{-}-\Sigma_{vc} ^{+}|^2} {(\epsilon-\Sigma_{cc} ^{+}-\Sigma_{vv}^{-})}.
\label{sec}
\end{equation}
To find the resonance solution at $\epsilon\sim\Delta_v$, note that the
smooth contributions $\Sigma_{ij}^+(\epsilon)$ (\ref{sigmaC})
can be neglected as compared
with the singular self energies $\Sigma_{ij}^-(\epsilon)$ (\ref{sigmaV}). 
The most singular term is estimated as 
$\Sigma_{cc}(\Delta_v)\sim \Delta_c/2$ (see upper panel of Fig. 5).  
Employing  an approximate value
of $\Sigma_{vv}^{-}(\Delta_v)\approx -\eta\Delta_c $ in the denominator of
the ratio on the r.h.s., the above equation reduces to,
\begin{equation}
2(\epsilon -\Delta) \approx 
\Sigma_{cc} ^{-}+\frac{\eta |\Sigma_{cc} ^{-}|^2}
{\Delta_v+\eta \Delta_c}\;.  
\label{seca}
\end{equation}
This equation has a Kondo-like pole at
$\epsilon=\Delta_v, T=\widetilde{T}_K$, where 
\begin{equation}
\widetilde{T}_K=D 
e^{-2\pi\Delta_c/\widetilde{\Gamma}_{c}},
\label{3.2}
\end{equation}
and $\widetilde{\Gamma}_{c}
\approx \Gamma_{c}(1-\eta\Delta_c/\Delta_v)$. 
As mentioned above, this pole does not imply the occurrence 
of a real bound state. It merely points out the characteristic 
temperature of crossover from the weak 
interaction regime to the strong coupling one.
The perturbation approach is valid only at $T>\widetilde{T}_K$. 

Beside the resonance at $\epsilon\sim \Delta_v$ there is of course
another resonance at $\epsilon\sim \Delta_c$. 
Repeating the above procedure, only the
terms $\Sigma_{ij}^+$ are retained in equation  (\ref{sec}),
and the midgap peak is found at $\epsilon=\Delta_c$, with 
characteristic temperature 
$\widetilde{T}^{'}_{K}=D\exp (-2\pi\Delta_v/\widetilde{\Gamma}_{v})$, 
with $\widetilde{\Gamma}_{v}\approx \Gamma_{v}\eta\Delta_c/\Delta_v$.
Since $\widetilde{T}^{'}_{K} < \widetilde{T}_K$, this peak is noticeably 
lower than the first one. 
The graphical solutions of equation (\ref{sec}) at
$T=\widetilde{T}_K$ are presented in the upper panel of Fig. 5.  

Thus, Kondo-type processes manifest themselves as a shake-up effect with
a shake-up energy $\Delta_{v,c}$. 
They can be related to a final state interaction
between the $(e,h)$ pair in the dot and the Fermi continuum in the leads. The
$T$-dependent logarithmic singularity in the exciton self energy is a
precursor of an "orthogonality catastrophe", 
in close analogy with the
corresponding anomaly encountered in connection with 
the self energy of a $d$-electron
within the 
conventional
Anderson model \cite{Lacr81}. 
In the latter case the Kondo peak transforms
to an undamped Abrikosov-Suhl resonance in the ground state
\cite{Glazr88,Gun83}.
However, this is not the case for the Kondo exciton 
(studied here) because of the finite
lifetime $\tau_l$ of the $(e,h)$ pair. The most important contributions to 
$\tau_l$ are given by the same tunneling processes which are responsible for
the very existence of the midgap states. To take them into account one
should include 
the states with $(e,h)$ pairs in the leads
in the Green function expansion. Consider first  
diagrams $(c)$ in Fig. 3, which describe the damping of a hole in 
the presence of an electron in the dot. 
Insertion of these diagrams in
the self energy $\Sigma_{cc}^-$ (\ref{1.15}) 
leads to the correction
\begin{equation}
B_{cc}^{-}  =  
\int^D_{-D}\frac{\bar{f}(\varepsilon^{\prime})d\varepsilon^{\prime}}
{\pi}\left[ \frac{\Gamma_{vv}}{\epsilon+\varepsilon^{\prime}-\varepsilon}+
\frac{2\Gamma_{cc}}{\epsilon+\varepsilon^{\prime}-\varepsilon+\Delta}\right].
\label{2.16a}
\end{equation}
Estimating these integrals near the singular point 
$\epsilon-\varepsilon_F\approx -\varepsilon_v$, we find that this correction 
results in an insignificant renormalization of the $v$-level position, 
that is,
\begin{equation}
\delta \varepsilon \sim \left(\Gamma_{cc}\ln \frac{|\varepsilon_f-\Delta|}
{D+\Delta}+\Gamma_{vv} \ln\frac{\Delta_v}{D}\right).
\label{ren}
\end{equation}
The valence level remains undamped in 
the case shown in fig. 4. In the general case of metallic lead with wide 
conduction band the damping is given by the diagram $(c)$ of Fig. 3 with
intermediate state "0" in a central loop (the first term in 
the r.h.s. of eq. \ref{2.16a}). The lifetime is determined by 
${\rm Im} B_{vv}^{-}\sim \Gamma_{vv}$, and the peak at $\epsilon\sim\Delta_v$ 
survives, provided $\Gamma_{vv}<\widetilde{T}_K$, i.e. 
$2\pi\Delta_c/\Gamma_{cc}<\ln D/\Gamma_{vv}.$ The last restriction is not
too rigid when the condition (\ref{3.0}) is valid.
 
Turning now to the second peak 
at $\epsilon\sim\Delta_c$ (Fig. 4), we see that 
its existence is limited by the damping effects which are contained in 
diagrams $(c)$ of Fig. 3. The corresponding correction to the
self 
energy $\Sigma^+_{vv}$ is given by the integral 
\begin{equation}
B_{vv}^{+}  =  
\int_{-D}^D\frac{f(\varepsilon^{\prime})d\varepsilon^{\prime}}{\pi}%
\left[ \frac{\Gamma_{cc}}{\epsilon -\varepsilon^{\prime}+\varepsilon}+ 
\frac{2\Gamma_{vv}}{\epsilon -\varepsilon^{\prime}+\varepsilon-\Delta}\right].
\label{2.16b}
\end{equation}
In this case the damping ${\rm Im} B_{cc}^{-}\sim \Gamma_{cc}$ is fatal
for the resonance because $\Gamma_{cc}>\widetilde{T}^{'}_{K}$. 

It has then been demonstrated above
that in the case of well localized hole states in the dot
the midgap  exciton arises at $\epsilon \sim \Delta_v$ provided the 
inequalities (\ref{3.0}) are satisfied. The crucial role of lifetime
effects
is especially manifested in the opposite limit of completely 
symmetric configuration $V_c=V_v$, $\Gamma_{ij}\equiv \Gamma$,  
$\Delta_c=\Delta_v \equiv \Delta/2
$. In this case $\Sigma _{0e}$ vanishes identically, and the secular
equation (\ref{2.10}) becomes
\begin{equation}
\epsilon -\Delta -\Sigma _{ee}(\epsilon )=0\; .  \label{2.17}
\end{equation}
Now ${\rm Re}\Sigma
_{ee}(\epsilon )$ diverges at $T\to \bar{T}_K=
D\exp \left( -\pi\Delta/ 2\Gamma \right).$
As a result a peak arises in ${\rm Im} G_{ee}$ at $\epsilon$ around $\Delta/2$ 
provided the lifetime effects are not taken into account. 
When the particle-hole symmetry is slightly violated ($\delta \neq
0$), $\Delta_{c} =\Delta /2-\delta $, $\Delta_v=\Delta /2+\delta $, $\delta
\ll \Delta /2$, this midgap peak disappears with 
increasing $\delta$ due to
cancellation of singular terms with opposite signs in eq. (\ref{2.10}). 
This midgap state is also fragile against the lifetime effect. 
In a symmetric case (\ref{2.17}) the self energies $B^{\pm}(\epsilon)$ have 
imaginary parts $\approx 2\Gamma$ at $\epsilon \approx \Delta/2$ 
which completely smear the peak structure. Thus, the
Kondo processes initiated by one of the partners in the electron-hole pair
are killed by the damping of its counterpart due to the same tunneling
mechanism. 

Focusing on a more promising asymmetric case (\ref{3.0}), we find that 
the main peak of optical transition at $h\nu = \widetilde{\Delta }$ 
is accompanied by a satellite peak at $h\nu \approx \Delta_v $. 
The form of this peak is determined by eq. (\ref{2.1}), i.e., by
\begin{equation}
\frac{1}{\pi}{\rm Im} G_{ee}(h\nu)= \frac{1}{\pi}{\rm Im} \left[ h\nu-\Delta-%
\widetilde{\Sigma}^-_{ee}(h\nu)+i\hbar\tau_l^{-1}\right] ^{-1},  
\label{2.19}
\end{equation}
where $\widetilde{\Sigma}^-_{ee}$ is given by the r.h.s. of eq. 
(\ref{seca}). The line-shape $W(h\nu)$ strongly depends on 
$T/\tilde{T}_K$ and $\tau_l$ \cite{Kibur93,Lacr81}. If the tunneling gives
no contribution to $\tau_l$, like in the case illustrated by fig. 3, 
the limiting factor is the recombination time of confined exciton in 
the dot. This time is measured, e.g., in nanosize Si
clusters embedded in amorphous SiO$_{2}$ matrix, and the experimentally 
estimated value is $\tau _{l}\sim 10^{-6}$ s for the singlet exciton 
\cite{Brong98}. Therefore, Kondo-type processes can survive in these 
systems if $\widetilde{T}_{K}\gg 10^{-9}$ eV.
Since $\widetilde{T}_K$ falls rapidly with increasing $\Delta_c$, this 
condition imposes limitations on the value of the ration 
$r \equiv 2\pi\Delta_c/\Gamma_{c}$: 
the shake-up sideband or satellite of the
exciton peak in luminescence spectra can be observed at 
$r< \ln(D/h\tau_l^{-1})$. Taking $D\sim 1 eV$, implies the restriction
$r<30$. Since the tunneling rate through the junction SQD/metal can be
made large enough (e.g., $\Gamma_{c} \sim 0.05$ eV for InAs/gold 
tunnel barrier \cite{Mil00}), this restriction is not insurmountable. 

Figure 5 illustrates the evolution of satellite peak in excitonic spectrum
with changing position of the level $\varepsilon_c$. The grafical solutions 
of eq. (\ref{sec}) for a Kondo temperature $\widetilde{T}_K$ are shown 
in the upper panel. Here the value of $\Gamma_{c}/2\pi=0.01D$ is chosen
and 
the hole life time is not taken into account. It is seen
from these solutions that the Kondo temperature rapidly falls from 
$\widetilde{T}_K\approx 1\cdot 10^{-7}D$ for $\Delta_c=0.15D$ (solid line) 
to $\widetilde{T}_K\approx 1\cdot 10^{-16}D$ for $\Delta_c=0.30D$ 
(dashed line). Due to the lifetime effects mentioned above the latter 
solution is inachievable, but the secondary shoulder or satellite of the
main excitonic peak arises in a first case. 
(lower panel of Fig. 5).  This panel illustrates the temperature
dependence of optical line-shape  
calculated by means of eq. ({\ref{2.19}). We have taken the 
value of $h\tau^{-1}_l=5\cdot10^{-9}D$ in these calculations. The shake-up 
satellites are seen distinctly at temperatures as high as 
$T\approx 10^2 \widetilde{T}_K$. 

It is worth stressing here that the exciton spectrum has been analyzed 
within the simplifying infinite-$U$ 
approximation. In particular,
it does not take into consideration charged 
states with two electrons or holes in the dot. Taking into
account finite $U$ means an inclusion of doubly occupied states
$\left|2c\right\rangle $, $\left| 2v\right\rangle $
in the set (\ref{1.6}), (\ref{1.2a}).
In close analogy with the procedure adopted for the
conventional Anderson model \cite{Gun83}, one
expects a redistribution of the spectral 
weight of neutral states $|0\rangle $
and $|es\rangle $ in favor of 
states $|2c\rangle $, $|2v\rangle $, and
an increment of $\widetilde{T}_{K}$ with decreasing $U$. However, the
inequality
$\widetilde{T}_{K}\ll \Delta_v $ ensures the existence
of the midgap states as discussed above.
 
\section{Further applications of the model}

When formulating the generalized Anderson tunnel Hamiltonian with 
electron-hole branches of excitations in a dot, 
the natural candidates for its realizations are
semiconductor nanoclusters. 
In this section some other physical systems for which
many-body satellites of exciton lines can arise are 
briefly discussed. It has been established
above that significant shake-up effects are expected
in the process of exciton transition provided that:
a) confined electron-hole
pairs form a discrete spectrum with 
large enough separation between the first
and subsequent levels, b) the Coulomb blockade is strong enough 
to satisfy the inequalities (\ref{1.3c}), 
and c) the hole states are only weakly 
damped. Evidently, SQD and wells containing 
transition metal (TM) impurities satisfy these conditions. 
TM impurities modify the electronic spectrum of semiconductor 
in such a way that the deep levels
appear in the energy gap \cite{KF94}. 
These levels are generated by  d-states
of unfilled 3d shell of TM ions which are substitution impurities  
characterized by the configuration $d^n$ 
in the ground (neutral) state. 
In spite of strong covalent effects
which mix the impurity $d$-states with Bloch waves 
pertaining to the host material bands (predominantly, with
valence $p$-states), these levels retain the $l=2$ 
angular symmetry modified 
by the cubic crystalline environment, 
and remain localized within 2-3 coordination 
spheres around the impurity site. Such TM impurity can bind an
electron-hole pair. Usually, either electron or hole appear in the 3d-shell,
and its counterpart is loosely bound by the Coulomb potential of ionized 
atom \cite{SK89}. Thus, the processes of electron-hole pair capture 
can be represented as reactions 
$d^n\to [d^{n-1}e]$ or $d^n\to [d^{n+1}h]$ ("donor" and "acceptor" exciton 
respectively). Both types of excitons are observed experimentally in 
bulk II-VI semiconductors doped by various TM impurities (see \cite{SK89}
for a review of experimental and theoretical results). 
The scheme of energy levels of a TM-bound donor exciton 
in a heterostructure with a band offset $\Delta\varepsilon_b$ is 
presented in Fig. 6 where the 
internal layer is selectively doped by 3d ions. 
Such heterostructure can be formed, e.g., in a three-layer sandwich 
CdSe/ZnTe/CdSe. The central layer should be doped by, 
e.g., Cr which is known to create a donor
exciton in CdSe \cite{KF94,SK89}. Here  
the deep level $\varepsilon_d$ 
is the energy of transition $d^n\to d^{n-1}$ 
in the impurity 3d shell.  
The shallow level near the bottom of the conduction band, 
$\varepsilon_{b}$, of the central layer is occupied by  an
electron bound to the positively charged $3d^{n-1}$ ion 
and {\it confined} in a 
barrier formed by the conduction band offset $\Delta\varepsilon_b.$ 
In order to prepare the conducting leads,
they should be heavily doped with $n$-type shallow impurities.

Since the states    
$d^{n}e$ and $d^{n-1}e^2$ are unstable due to Coulomb repulsion effect,
and all states including $d^{n-2}$ configuration are highly excited on 
an atomic scale, the pertinent physics is 
faithfully encoded by the model Hamiltonian 
(\ref{1.1}) where the term $H_d$ (\ref{1.3d}) includes the donor exciton 
described above. Apparently, the wavefunctions of  electrons in the
d-shell do not overlap with the
band electron states in the leads. On the other hand, 
these band electrons can tunnel between the leads through the resonance 
shallow level $\varepsilon_c$ of the confined electron. 
Hence, the properties of this system correspond to the case
discussed at the beginning of Section IV, and one can expect 
the appearance of a satellite exciton 
at an energy $\Delta_V=\varepsilon_{F}-\varepsilon_d$.  

Another possible realization of our model 
is exposed when mixed valent rare earth atoms are adsorbed on
a metallic surface (hereafter they are referred to as adatoms). 
It is known that the Anderson model can be applied to
adatoms with strongly interacting electrons (see\cite{Grim} for a review).
In this case $H_{t}$ (\ref{1.2f}) corresponds to covalent bonding between an
adatom and a substrate, $V$ is the corresponding hybridization
integral between the electrons in the adatom and those in the nearest sites of
the metallic surface layer, 
and $U$ is the intra-atomic Coulomb repulsion which
prevents charging of the adatom in the process of chemisorption. The model
was originally proposed for hydrogen atoms adsorbed on surfaces of
transition metals \cite{New69}. Later on, the possibility of Kondo-type
spin polarization of substrate electrons around 
an adatom spin in the case of $%
U\gg V^{2}/D$ was discussed \cite{Schrief74}. The most promising candidates
from the point of view of exciton effects are adatoms with unstable
valence, e.g., Sm, whose ground state electronic configuration is $%
4f^{6}6s^{2}$. Sm atoms can indeed 
be adsorbed on surfaces of transition metals such as
Ni, Co, Cu, Mo. In the process of adsorption, 
a Sm atom loses its $s$-electrons
and exists in two charged states Sm$^{2+}$ and Sm$^{3+}$ depending on the
concentration of Sm ions on the surface. In particular, isolated ions Sm$%
^{2+}(4f^{6})$ are observed on Mo surface at low submonolayer coverage \cite
{Sten89}. The unfilled 4f shell forms a resonant f-state close to the Fermi
level of the metal. The excited 5d state forms another level above $%
\varepsilon _{F}$. Thus, one 
arrives again at a two-level system described by the
Hamiltonian $H_{d}$ (\ref{1.2}). Since the ground state term of the
configuration $4f^{6}$ is a singlet $^{7}F_{0}$, one cannot expect 
the occurrence of Kondo
coupling for such adatom. However, in the course of virtual transitions
between the adatom and the substrate the states $|kv\rangle $ and $%
|kc\rangle $ appear with excess electron $e_k$ above $\varepsilon_F$
(configuration $4f^{5}e_k$) and a hole $h_k$ 
below $\varepsilon_F$ (configuration $4f^{5}5dh_k$). 
According to our calculations, one can excite not only the
conventional atomic excited state with energy $\Delta =E(4f^{5}5d)-E(4f^{6})$
but also the midgap states with energy close to $\Delta ^{\prime
}=E(4f^{5}e_{F})-E(4f^{6})$ where $e_{F}$ refers to electrons on the
Fermi level of the substrate.

\section{Conclusions}

To summarize, this work suggests a generalization of the Anderson impurity
model for semiconductor quantum dots in contact with metallic leads. 
This model takes into account the exciton 
degrees of freedom which are absent in a standard theory, and the spectrum 
of neutral excitation involves not only the electron-hole pairs confined
within the dot, but also states 
in which one of the carriers appears in the 
leads in a process of tunneling. As a result, the model exhibits 
precursors of a Kondo effect 
pertaining to an excited state of an even-${\cal N}$ SQD, 
despite the  absence of Kondo coupling in the ground state. 
The theory developed here has an immediate experimental
prediction, namely, that satellite exciton peaks of a Kondo
origin can be detected in the 
optical absorption or luminescence spectra 
of SQD.

Attention in this paper was focused
on the optical properties of SQD, whereas the physics of
tunneling transport through these dots 
is expected to manifest
specific features as well. 
Evidently, occurrence of midgap resonances of exciton type can
not  noticeably influence the single-electron or single-hole tunnel 
conductance of SQD, but anomalies in two-particle transport can 
exist. Novel
features might be expected in luminescence relaxation 
spectra and the corresponding  photocurrent. 
Moreover, an anomalous 
energy transport due to electron-hole cotunneling through SQD is also 
possible. In particular, 
investigation of frequency dependent photocurrent 
seems to be our next objective.

It is useful to point out once more that 
the case studied here refers to SQD with wide forbidden gap.
The energy of excited neutral dot is higher than the energy of a singly 
charged dot. Such SQDs exist in a form of nanosize semiconductor droplets.
Another type of quantum dots which is created near heterostructure
interfaces by means of external gate voltage is an example of the 
opposite limit of complete Coulomb blockade, when the charging energy is
much higher than the energy of the confined exciton. Our studies show that
unusual Kondo type effect are possible in both cases. The difference
is that in the first case, these effects appear as shake-up
processes in excited states, and manifestations of many-body 
resonances are limited by 
the finite life time of the confined excitons, whereas in the second case 
a Kondo-type ground state can emerge provided triplet excitons
are involved in the tunneling processes, and hence, spin-flip transitions
become possible. 

\noindent
{\bf Acknowledgment}: 
We are very much indebted to M. Pustilnik who participated in 
most of our pertinent discussions. His comments, 
suggestions and criticisms were invaluable for completing this 
research. 
We thank A. Polman, M. Brongersma and O. Millo 
for stimulating discussions of the experimental properties of 
semiconductor nanoclusters. The valuable assistance of I. Kikoin in 
numerical calculations is gratefully acknowledged. This
research is supported by
the Israel Science Foundation grants 
"Nonlinear Current Response of Multilevel Quantum
Systems" and "Strongly Correlated Electron Systems in Restricted Geometries", 
by DIP program ``Quantum electronics in low dimensional systems'' and a 
BSF program ``Dynamical instabilities in quantum dots''.

\newpage
{\Large {\bf Figure Caption}}
\bigskip\\ 
{\large Fig. 1}. Building blocks of the Green function
expansion and the secular equation (\ref{2.10}). The arrows indicate the
tunneling processes which connect different states of this set.
\bigskip\\
{\large Fig. 2}. Vertices, which couple the ground state 
$\langle 0|$ and the singlet exciton $\langle es|$ with the electron-hole
pair states $|kc\rangle,|kv\rangle$
\bigskip\\
{\large Fig. 3} Second and fourth order 
diagrams for singlet exciton self energy
$\Sigma_{ee}$ 
\bigskip\\
{\large Fig. 4}. Energy levels of SQD coupled with degenerate $n$-type 
semiconductor
\bigskip\\
{\large Fig. 5}. Upper panel: Graphic representation of 
eq. \ref{sec} at $T =\widetilde{T}_K$.
$\widetilde{\Sigma}_{ee}$ is the right hand side of eq. \ref{sec}.
$\Delta_c=0.15D$, $\Delta_v=0.85D$ (solid line), 
$\Delta_c=0.3D$, $\Delta_v=0.7D$ (dashed line).
Lower panel: Optical line-shape $W$ calculated from eq. (\ref{2.19}) 
for $\Delta_c=0.15D$. $T=10^2 \widetilde{T}_K$ (dashed line), $T=5\cdot
10^2 \widetilde{T}_K$ 
(solid line). $W_0$ is the maximum value of $W$.
\bigskip\\
{\large Fig. 6}. Structure of electron bands and bound exciton states in 
a heterostructure formed by a central layer of semiconductor doped 
by TM impurities and the left and right layers formed by n-doped 
semiconductors.   
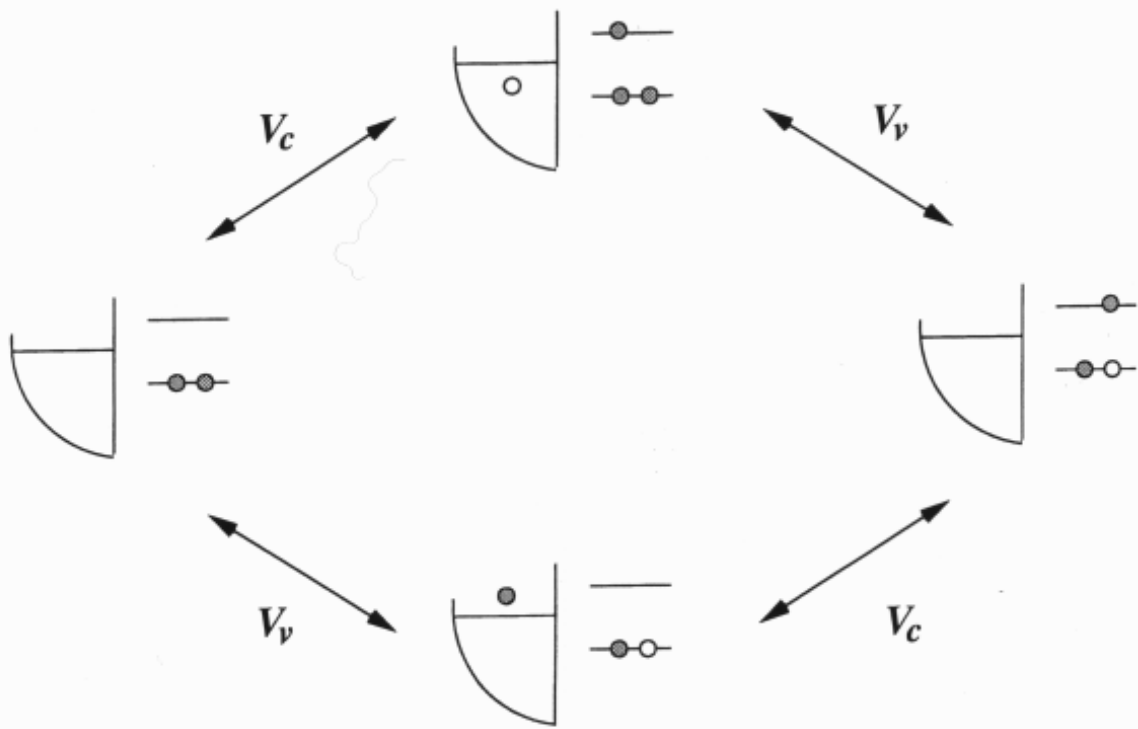
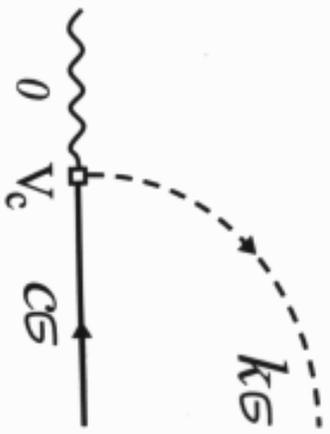
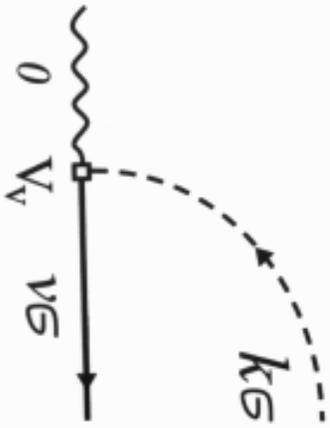
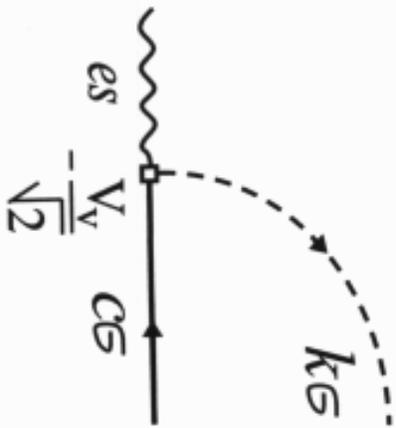
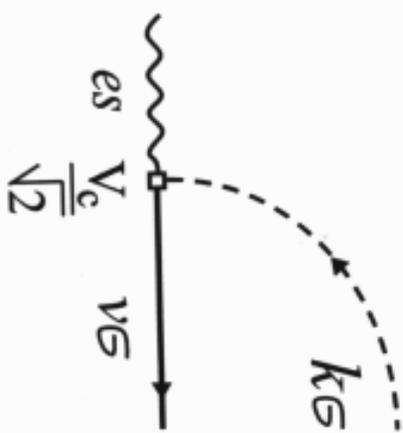
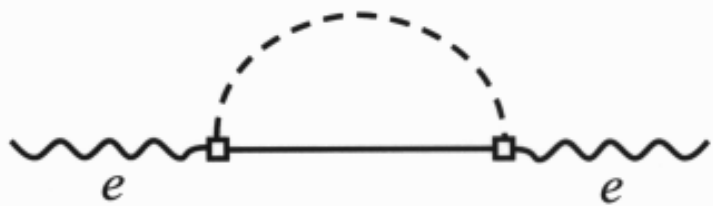
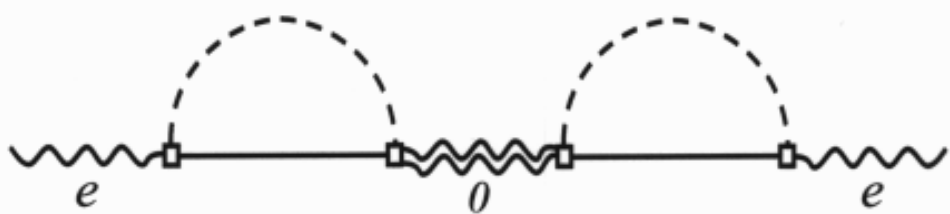
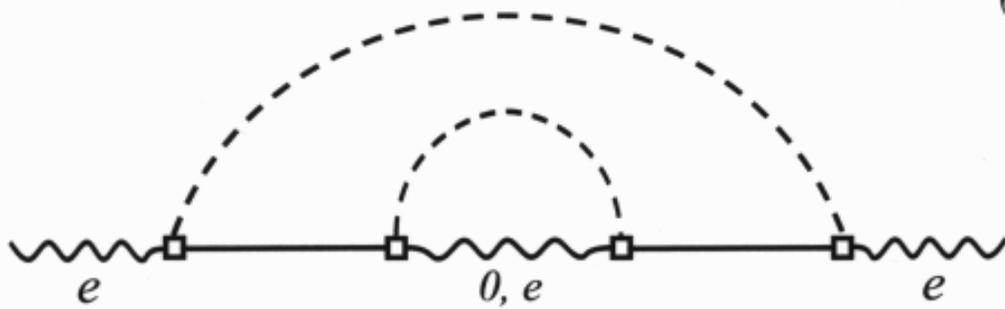
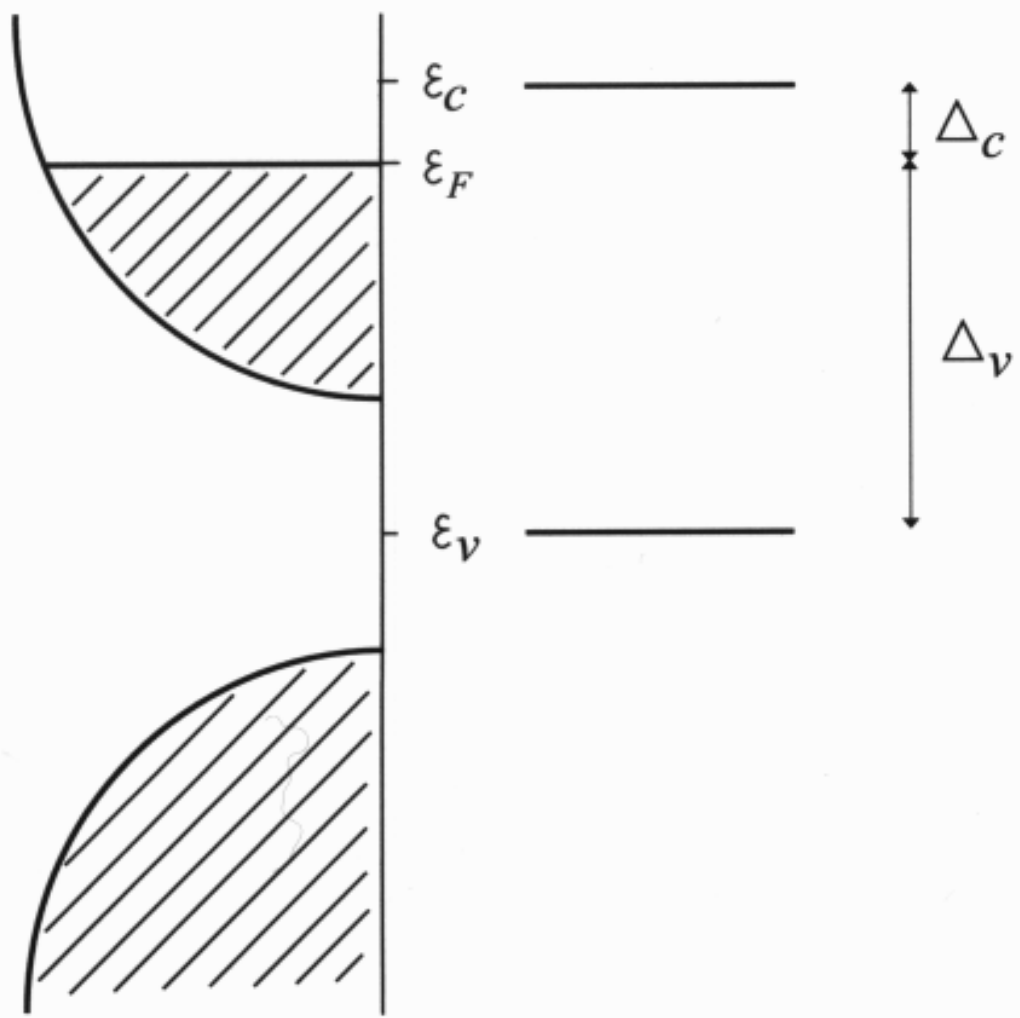
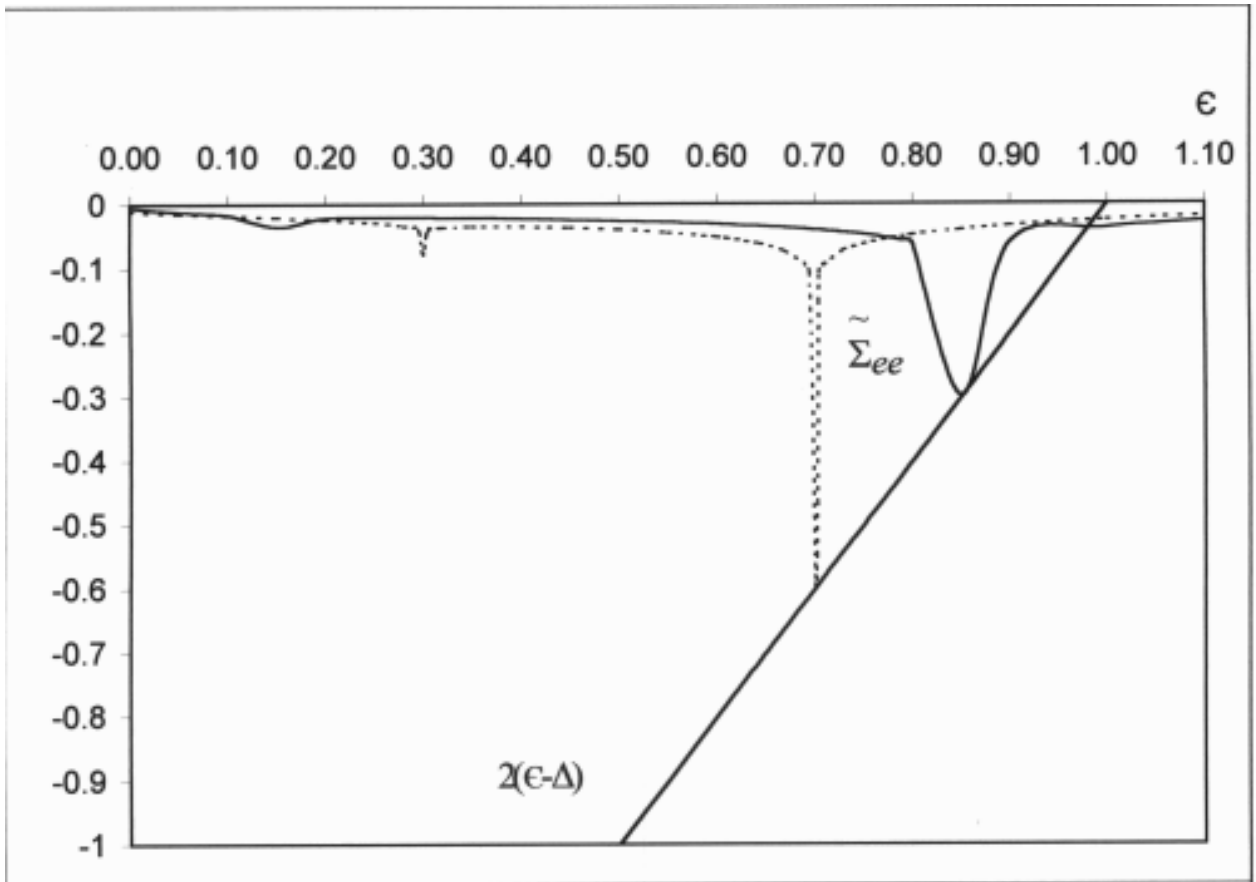
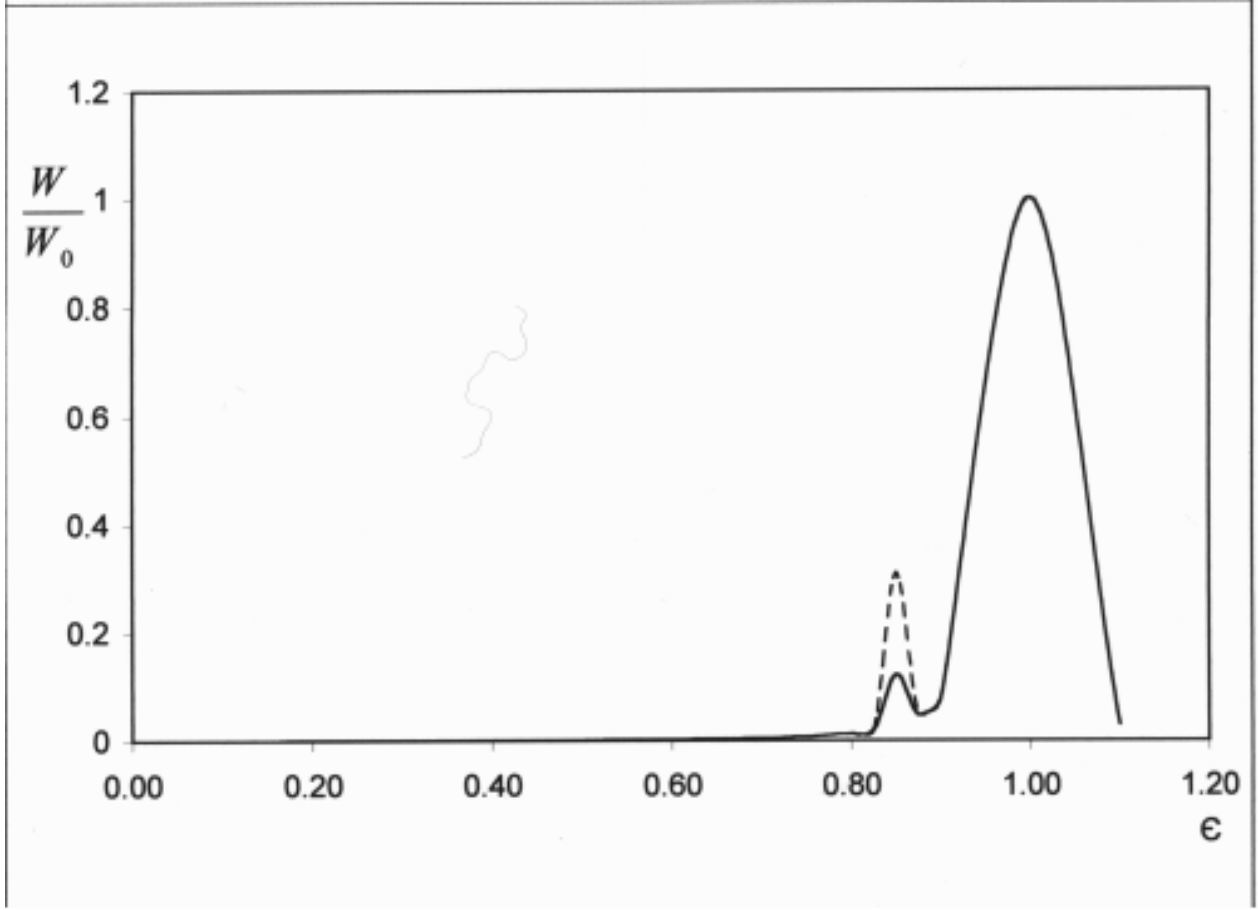
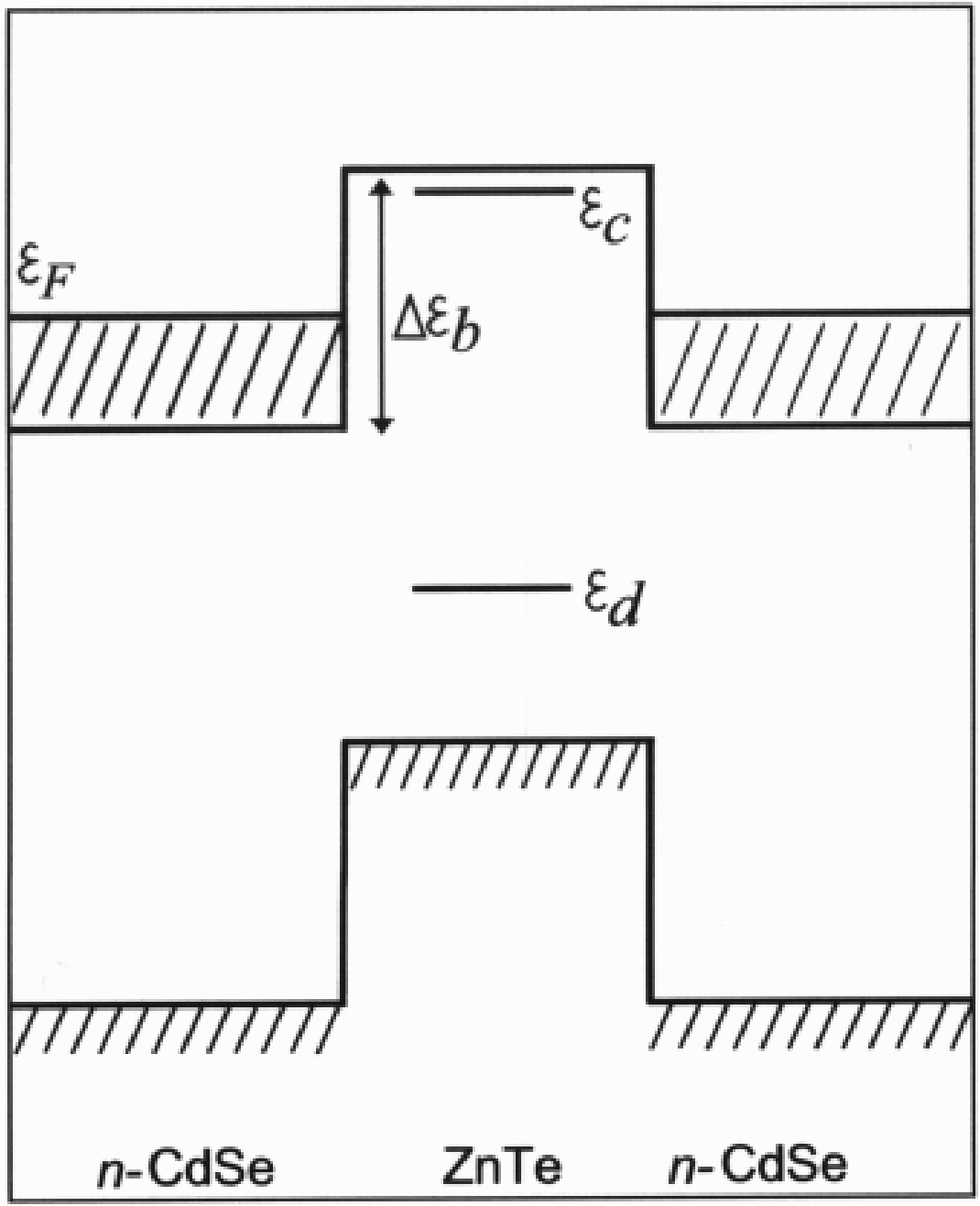
\end{document}